\documentclass[fleqn,usenatbib]{mnras}

\usepackage{newtxtext,newtxmath}

\usepackage[T1]{fontenc}

\DeclareRobustCommand{\VAN}[3]{#2}
\let\VANthebibliography\thebibliography
\def\thebibliography{\DeclareRobustCommand{\VAN}[3]{##3}\VANthebibliography}

\usepackage{graphicx}	
\usepackage{amsmath}	

\usepackage[utf8]{inputenc} 
\usepackage[T1]{fontenc}    
\usepackage{hyperref}       
\usepackage{url}            
\usepackage{booktabs}       
\usepackage{amsfonts}       
\usepackage{nicefrac}       
\usepackage{microtype}      
\usepackage{xcolor}         

\usepackage{mathtools}
\usepackage{afterpage}

\title{Virgo: Scalable Unsupervised Classification of Cosmological Shock Waves}

\author[M. Lamparth et al.]{
Max Lamparth,$^{1, 3}$\thanks{Corresponding author: max.lamparth@tum.de}
Ludwig M. B\"oss,$^{2, 3}$
Ulrich P. Steinwandel$^{4}$
and Klaus Dolag$^{2, 3, 5}$ \\
\\
$^{1}$Physik-Department, Technische Universit\"at M\"unchen, James-Franck-Str. 1, Garching, 85748, Germany \\
$^{2}$Universit\"ats-Sternwarte, Fakult\"at für Physik, Ludwig-Maximilians-Universit\"at M\"unchen, Scheinerstr.1, M\"unchen, 81679, Germany \\
$^{3}$Excellence Cluster ORIGINS, Boltzmannstr. 2, Garching, 85748, Germany \\
$^{4}$Center for Computational Astrophysics, Flatiron Institute, 162 5th Avenue, New York, 10010, United States of America \\
$^{5}$Max-Planck-Institut für Astrophysik, Karl-Schwarzschild-Str. 1, Garching, 85741, Germany
}

\date{Accepted XXX. Received YYY; in original form ZZZ}

\pubyear{2022}

\begin{document}
\label{firstpage}
\pagerange{\pageref{firstpage}--\pageref{lastpage}}
\maketitle

\begin{abstract}
    Cosmological shock waves are essential to understanding the formation of cosmological structures.
    To study them, scientists run computationally expensive high-resolution 3D hydrodynamic simulations.
    Interpreting the simulation results is challenging because the resulting data sets are enormous, and the shock wave surfaces are hard to separate and classify due to their complex morphologies and multiple shock fronts intersecting.
    We introduce a novel pipeline, \textsc{Virgo}, combining physical motivation, scalability, and probabilistic robustness to tackle this unsolved unsupervised classification problem.
    To this end, we employ kernel principal component analysis with low-rank matrix approximations to denoise data sets of shocked particles and create labeled subsets.
    We perform supervised classification to recover full data resolution with stochastic variational deep kernel learning.
    We evaluate on three state-of-the-art data sets with varying complexity and achieve good results.
    The proposed pipeline runs automatically, has only a few hyperparameters, and performs well on all tested data sets.
    Our results are promising for large-scale applications, and we highlight now enabled future scientific work.\\
\end{abstract}





\section{Introduction}
\label{sec:intro}


 
\begin{figure*}
    \centering
    \includegraphics[width=\textwidth]{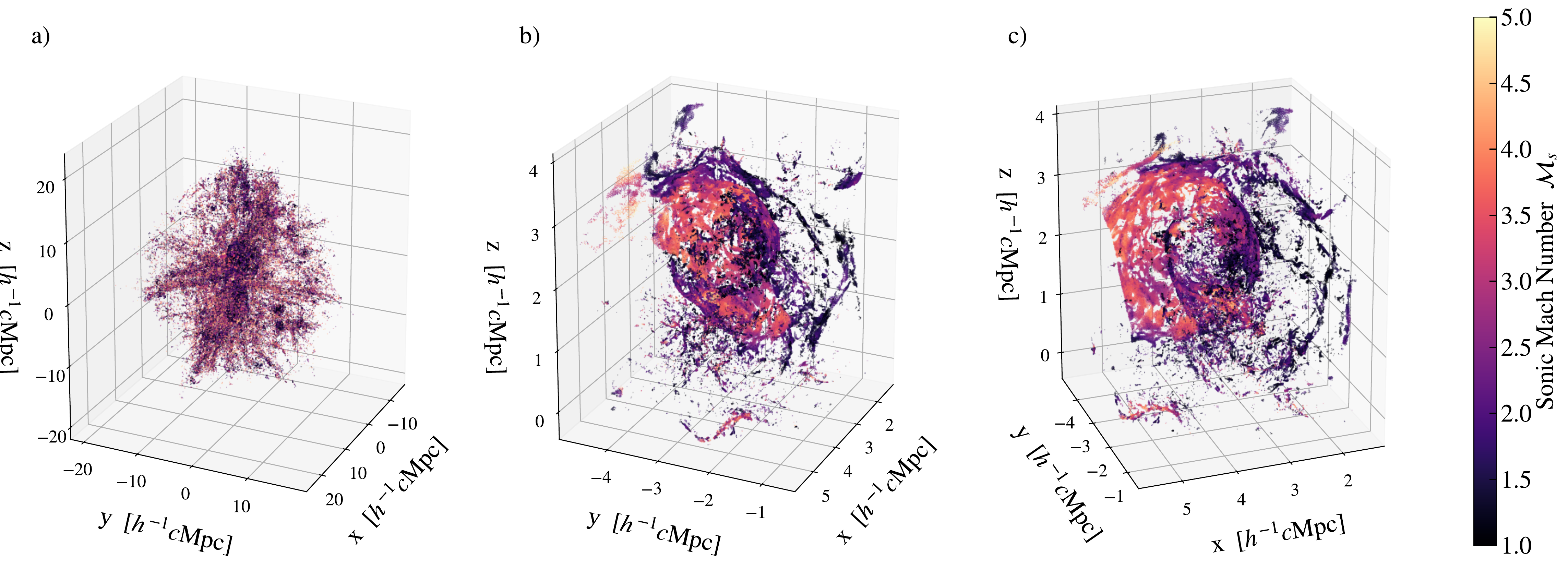}
    \caption{We show a reduced simulated data set $\textsc{ClustHD}_2$ containing all particles with a detected shock in the sonic Mach number range $\mathcal{M}_s \in [1, 5]$. \textit{a)} Full simulation domain. The main cluster is in a cutout of the cosmic web, which shows internal and accretion shocks. \textit{b)} Manual zoom into the cluster showing the complex shock structure of multiple ongoing merger shocks. We want to determine all shock surface particles as separate labeled groups and remove non-shock wave particles. \textit{c)} Same as \textit{b)}, but rotated by 45$^\circ$.}
    \label{fig:raw250x_shocks}
\end{figure*}

Cosmological structures form by gravitationally accreting mass from their surroundings \citep[e.g.][]{Bertschinger1998, Somerville2015, Naab2017}.
As galaxies and groups of galaxies fall into clusters, they dissipate their energy in the form of shock waves in the diffuse gas between them, labeled as the intra-cluster medium (ICM) \citep[e.g.][]{Dolag2005, Pfrommer2007, Pfrommer2016, Vazza2011, Vazza2016, Schaal2015, Steinwandel2021}.
Within these systems, the evolution of shock waves is the primary driver that sets the global physical properties of these systems, like the velocity and temperature structure.
In addition, shock waves can play a crucial role in the evolution of many other astrophysical systems such as supernova explosions in the Interstellar medium \citep[e.g.][]{Sedov1946, Taylor1950a, Taylor1950b, Kim2015, Walch2015, Steinwandel2020, Fielding2017}. 
These shock waves are defined as discontinuities in density and temperature, propagating through the hot ($T \sim 10^8 \mathrm{K}$) and thin ($\rho \sim 10^{-28} \mathrm{g} \: \mathrm{cm}^{-3}$) plasma of the ICM at supersonic speeds.
They are among the universe's most powerful accelerators of relativistic protons and electrons (cosmic rays, CRs).
Moreover, we can observe them as arc-shaped radio synchrotron emission sources on the outskirts of merging galaxy clusters, millions of lightyears in diameter \citep[][]{Weeren2019}.
The total energy dissipated at these shocks which will subsequently become available for accelerating CR is proportional to the shock speed and surface.
As the fraction of shock wave energy available to accelerate CRs is still poorly constrained \citep[][]{Kang2007, Kang2013, Guo2014, Caprioli2014, Caprioli2018, Ryu2019, Ha2021}, knowing the total energy budget helps to constrain the theoretical predictions of acceleration efficiency and therefore give preferences to the different formation scenarios. 
A statistical comparison of large-scale cosmological simulations with large observational campaigns has the potential to better discriminate between the different formation scenarios. 
Modeling these cosmological systems with state-of-the-art simulations requires $\mathcal{O}(10^{6})$ CPU hours and $\mathcal{O}(10^{2})$ TB of RAM, only available on modern supercomputers, as there is large degeneracy in the possible geometry and underlying model assumptions.
The produced data sets contain up to $\mathcal{O}(10^{10})$ particles, which we need to interpret to make conclusions about formation scenarios.
However, shock wave structures in galaxy clusters form highly complex shapes and surfaces (see Fig.~\ref{fig:raw250x_shocks}), and collisions between them can lead to a complex superposition of different shock waves with overlapping geometries. Therefore we can not make a simple, prior connection between in-falling substructure as drivers with the associated parts of the shock surface from first principles.
This setup poses a complex unsupervised classification problem for an unknown number of target classes in which we must find, separate, and label coherent shock wave structures in simulated data. \\

To this end, we propose a novel, physically motivated, and fully scalable machine learning pipeline to solve the outlined unsupervised classification problem. 
We separate the task and classify a subset of each original unlabeled data set. 
For the subset classification, we exploit the inherent non-stationarity of the problem with kernel functions for data pre-selection and use Gaussian mixture models \citep{barber2012bayesianml} to pre-clean the data from unwanted non-shock wave particles.
We further use physically motivated kernel functions with kernel principal component analysis \citep{kernelpca_scholkopf1998} and low-rank matrix approximations \citep{nystrom_drineas2005, GpRasWil} to map the data set to a favorable feature space.
In this feature space, the data is separable with stationary algorithms as the mapping reduces the local density changes of the original data set.
To emphasize this, we use a simple agglomerative clustering algorithm with an automatically determined linking length.
Finally, we use the labeled subset to train a scalable classifier to use our algorithm on the full data sets.
The classifier needs to be able to deal with the complex density changes of the data set.
Therefore, we combine the flexibility of the adaptive basis functions of a neural network with the infinite set of fixed basis functions of a Gaussian process kernel in a scalable way for stochastic variational deep kernel learning (SV-DKL, DKL), as proposed by the authors of \citep{svgp_hensman2013, svgpclass_hensman2015, dkl_wilson2016, svdkl_wilson2016}. 
For the first time, we can tackle this previously unsolved problem with our described pipeline and guarantee scalability for current state-of-the-art and future data sets.
To the best of our knowledge, this is also the first application of DKL and SV-DKL to an astrophysical task.


\section{Background and Related Work}
\label{sec:background}


For our work, we focus on distribution-based (Gaussian mixture models (GMM) \citep{barber2012bayesianml}), connectivity-based (hierarchical clustering, friends-of-friends (FoF) \citep{fof_davis1985}), and density-based spatial clustering (DBSCAN \citep{dbscan_ester1996}) algorithms.
Our problem requires a flexible solution that makes little or no prior assumptions about shapes and the number of target classes and can deal with non-euclidean metrics.
We focus on the listed approaches as they offer various degrees of good scalability to large data sets \citep{mahdi2021scalable} and flexibility.
The FoF algorithm is a special case of DBSCAN \citep{kwon2010scalable} and is used in astrophysics, e.g., for structure identification \citep{fof_davis1985, Dolag2009_subfind}. 
GMM without a Dirichlet prior \citep{barber2012bayesianml} require prior assumptions about the expected number of classes.
Another tool we can use for dimensionality reduction in combination with previously stated algorithms is principal component analysis (PCA) \citep{pca_jolliffe2016}.

\subsection{Kernel Methods}

Kernel functions $k$ are defined as integral operators. Should $k$ fulfill the Mercer condition \citep{kernel_scholkopf2018}, calculating $k(\mathbf{x}, \mathbf{x}^\prime)$ is equivalent to mapping $\mathbf{x}$ and $\mathbf{x}^\prime$ to a feature space via a function $\phi$ and taking the inner product \citep{kernel_scholkopf2018}
 \begin{align}
     k(\mathbf{x}, \mathbf{x}^\prime) = \langle \phi(\mathbf{x}), \phi(\mathbf{x}^\prime) \rangle_{\nu} .
 \end{align}
This equivalence is called the \textit{kernel trick}.
Since the equivalent mapping of $\phi$ is primarily non-linear and usually into higher dimensions, we can use kernel functions to make data more separable for a more straightforward classification.
Furthermore, we can construct the Gram matrix $\mathbf{K}$ for $K_{ij} = k(\mathbf{x_i}, \mathbf{x_j}^\prime)$ with respect to a data set $\mathcal{D}_n = \{ \mathbf{x_1}, ..., \mathbf{x_n} \}$ of size $n$.
Should $\mathbf{K}$ be positive-semi definite (PSD), then $k$ is a valid covariance function and $\mathbf{K}$ the corresponding covariance matrix.
We can interpret $\mathbf{K}$ as a similarity matrix because we score each point $\mathbf{x_i}$ to each other point in $\mathcal{D}_n$ by a similarity criterion defined by the kernel function $k$.\\
\\
Kernel functions can be stationary, i.e. $k(\mathbf{x}, \mathbf{x}^\prime) = k ( \mathbf{x} - \mathbf{x}^\prime )$, and therefore invariant under spatial displacements. 
In addition, we can create composite kernels, as a sum, product, or convolution of two kernel functions \citep{kernel_scholkopf2018}.
Given that $k$ scores points by a similarity criterion, we can interpret composite kernels as combining similarity criteria with logical operators.
As the kernel function parameters are related to data-specific characteristics such as length scales, kernel methods offer interpretable parameters. This benefit is unique compared to other machine learning methods and makes them attractive for physics applications \citep{gp_lamparth2022}. 
Thus, we can create physically motivated kernel functions with physical parameters and encode physical information like, e.g., symmetry or local density changes of our problem. \\
\\
The computational complexity of $\mathbf{K}$ scales with $\mathcal{O}(n^2)$.
To use kernel functions with large data sets, we can use a low-rank matrix approximation called Nyström approximation \citep{nystrom_drineas2005, GpRasWil}.
For this, we chose a random subset $\mathcal{D}_m$ of $m$ data points with $ m < n$ to represent the data set and calculate the Gram matrix for $K_{ij} = k(\mathbf{x_i}, \mathbf{x_j})$ with $\mathbf{x_i} \in \mathcal{D}_n = \{ \mathbf{x_1}, ..., \mathbf{x_n} \}$ and $\mathbf{x_j} \in \mathcal{D}_m \subset \mathcal{D}_n$.
Furthermore, we can expand regular PCA to non-linear component analysis by constructing $\mathbf{K}$ of our data set and performing PCA on the result \citep{kernelpca_scholkopf1998}.

\subsection{Gaussian Processes and Stochastic Variational Gaussian Processes}

Gaussian processes provide a strong mathematical framework for probabilistic and non-parametric regression and classification \citep{GpRasWil}. 
For $n$ observations $\mathbf{y}$ at input locations $X$, we can use a PSD kernel function and its Gram matrix $\mathbf{K}$ as a covariance matrix and optimize the model parameters by maximizing the marginal log-likelihood
\begin{align}
    \log p(\mathbf{y} \mid X) &= -\frac{1}{2} \mathbf{y}^{\top} \mathbf{K}^{-1} \mathbf{y} -\frac{1}{2} \log \left|\mathbf{K}\right|-\frac{n}{2} \log 2 \pi .
    \label{equ:mll}
\end{align}
Due to the inversion of $\mathbf{K}$ in Equ.~\eqref{equ:mll}, Gaussian processes have a computational complexity of $\mathcal{O}(n^3)$.
For classification, the likelihood is non-Gaussian, intractable and therefore needs to be approximated \citep{GpRasWil}.
An approach to increasing Gaussian processes' scalability are stochastic variational Gaussian processes (SVGPs), as outlined by \citep{svgp_hensman2013, svgpclass_hensman2015, mcmc_hensman2015}. 
They combine sparse Gaussian processes \citep{sgp_candela2005} with posterior approximation via variational inference \citep{vi_blei2017}.
In addition, the approach uses $m$ trainable inducing points $X_{S}$ with $m < n$, to have $X_{S}$ and their predictions $\mathbf{f}_S$ summarize the data set. 
In doing so, the authors enable non-Gaussian likelihoods and stochastic gradient descent optimization on a lower bound of the marginal log-likelihood from Equ.~\eqref{equ:mll}.
SVGPs achieve a computational complexity of $\mathcal{O}(m^3)$.
However, SVGPs are limited to problems where a smaller number of inducing points suffice to represent the data set's details and where regular kernels are expressive enough.

\subsection{Stochastic Variational Deep Kernel Learning}
\label{subsec:svdkl}

To overcome the limits of kernel functions, the authors of \citep{dkl_wilson2016} introduced scalable deep kernel learning (DKL) to utilize the adaptive basis functions of a neural network (NN).
To achieve computational complexity of $\mathcal{O}(n)$ for $n$ training points during inference, they use local kernel interpolation, inducing points, and structure exploiting algebra introduced in \citep{kissgp_wilson2015}.
The authors of \citep{dkl_wilson2016} use the output features of a deep kernel NN as input for a base kernel of a Gaussian process and jointly train the parameters of the deep kernel NN and the base kernel.
This approach combines the, e.g., infinite set of fixed basis functions of a radial basis function (RBF) kernel with a set length scale with a fixed set of highly adaptive basis functions of NNs.
Thus, the deep kernel NN learns statistical representations from the data set and enables non-euclidean similarity metrics throughout the input space. \\
\\
The DKL approach was expanded by \citep{svdkl_wilson2016}, introducing stochastic variational deep kernel learning (SV-DKL).
It allows for stochastic gradient training enabled by variational inference \citep{vi_blei2017} through a marginal log-likelihood lower bound objective.
For correlated predictions, the authors feed the outputs of the deep kernel NN into a set of additive base kernels, which they use for independent Gaussian processes.
Furthermore, they mix these Gaussian processes linearly to induce correlations and feed the outputs into a softmax likelihood for classification.
The SV-DKL model achieves a computational complexity of $\mathcal{O}(m^{1+1/d})$ for fixed inducing points $m$ and input dimensions of our data set $d$.
We apply SV-DKL on data sets with $d = 6$, see Sec.~\ref{sec:virgo}.

\subsection{Related Astrophysical Applications}

Gaussian processes have found applications in astrophysics \citep{gpcosmo_shafieloo2012, gppulsars_haastere2014, gpgravwaves_moore2016, bayoptcosmo_leclercq2018} as well as (S)VGP \citep{astroGP_karchev2021, svgpdust_miller2022}.
The authors of \citep{gp_lamparth2022} suggested SVGP and SV-DKL for large-scale physics simulation interpolation.
There have been analytical \citep{analshock_snow2021} and deep learning \citep{nn4shocks_morgan2020, deepzipper_morgan2022} approaches to shock wave and supernovae identification, whereas only \citep{analshock_snow2021} did classification.
The authors of \citep{svm_wang2022} recently applied support vector machines to the supervised classification of stellar objects.  
Clustering algorithms are commonly used in astrophysics \citep{fof_davis1985, autoclustering_zhang2004, discfinder_fu2010, asteca_perren2015, pyupmask_pera2021}.
Compared to the other outlined models, SV-DKL is the only approach offering Gaussian process non-parametric and statistical benefits, sufficient scalability and flexibility of applied metrics to solve our problem.
To the best of our knowledge, this is the first application of DKL and SV-DKL to an astrophysical task.
Previous work only hinted at DKL applications \citep{BayAttNP_park2021}.
Furthermore, we are not aware of any preceding work having solved this unsupervised classification problem, in particular not with state-of-the-art resolution data sets.

\section{Data Sets}
\label{sec:datasets}

\begin{figure*}
    \centering
    \includegraphics[width=\textwidth]{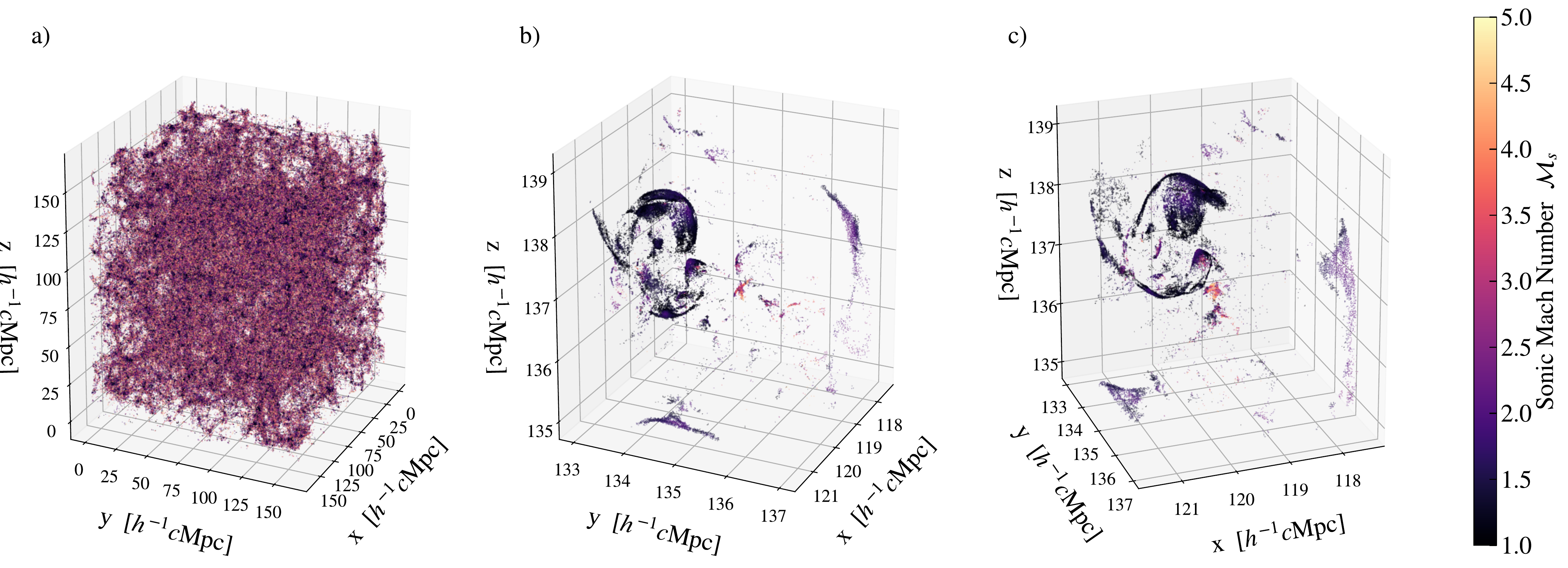}
    \caption{We show a reduced dataset for the \textsc{BoxMHD} simulation, as in Fig.~\ref{fig:raw250x_shocks}.  $a)$ Full simulation domain. As this is a cosmological box simulation it contains the full cosmic web. \textit{b)} Manual zoom on one of the galaxy clusters forming at the intersection of cosmic filaments. \textit{c)} Same as \textit{b)}, but rotated by 45$^\circ$.}
    \label{fig:25x_box_shocks}
\end{figure*}

We employ three data sets for the shock surface classification to gauge the flexibility of our machine learning pipeline for increasing physical complexity and different scales.
The data sets we are using do not contain any personally identifiable information or offensive content.\\
\\
All data sets were generated with \textsc{OpenGadget3}, which is an advanced version of \textsc{P-Gadget3} \citep{Springel2005} with an updated SPH scheme \citep{2016MNRAS.455.2110B}. \textsc{Gadget} computes gravity via a tree algorithm \citep{Barnes1986} and solves for hydrodynamics using smoothed particle hydrodynamics \citep[SPH,][]{Lucy1977, Gingold1977}.
Additionally, we adopt numerical schemes for thermal conduction \citep[e.g.][]{Jubelgas2004, Arth2014}, on-the-fly shock detection \citep[][]{Beck2016a}
and a non-ideal magneto-hydrodynamics (MHD) scheme \citep[][]{Dolag2009,Bonafede2011}.
Different time steps of the same simulation are quasi-independent data sets to be solved due to changing shapes, morphing structures, and a changing number of target classes.
We therefore use different time steps of the same simulation as different data sets and indicate them with an index, e.g., during experiments in Sec.~\ref{sec:appl}.
We name the data sets as follows:
\begin{itemize}
    \item \textsc{ClustHD}: Ultra-high-resolution hydrodynamics simulation of a single massive galaxy cluster, as shown in Fig.~\ref{fig:raw250x_shocks}. 
    \item \textsc{ClustMHD}: Same as \textsc{ClustHD}, but expanded to magneto hydrodynamics.
    \item \textsc{BoxMHD}: High-resolution simulation of a large magneto hydrodynamical cosmological volume with many clusters, as shown in Fig.~\ref{fig:25x_box_shocks}. 
\end{itemize}
\textsc{ClustHD} and \textsc{ClustMHD} are re-simulations in the cosmological simulation \textsc{COMPASS} and \textsc{Dianoga} \citep[for a detailed description of the setup see][]{Bonafede2011} with $\sim 5 \cdot 10^{8}$ particles\footnote{You can find more information, as well as a movie of the simulation at \url{http://www.magneticum.org/complements.html##Compass}}.
\textsc{ClustHD} has recently been used to study the interaction of shock waves in the ICM \citep[][]{Zhang2020}.
As \textsc{ClustMHD} also contains magnetic fields, which can lead to additional random motions perpendicular to the shock propagation and a more patchy shock wave surface, it is an ideal case to extend the underlying physical model to probe the robustness of our method.
%
The third data set is a cosmological volume simulation with a box size of 240 Mpc and an effective mass resolution of $\sim 4.5 \cdot 10^{6}$ M$_{\odot}$ which corresponds to $2048^3$ particles in the box for the dark matter component and the gas component respectively. 
We adopt a Planck-cosmology \citep[][]{Planck2016} with $\Omega_\mathrm{m} = 0.308$, $\Omega_{\Lambda} = 0.692$, $\Omega_{b} = 0.049$, $H_{0} = 67.81$ km s$^{-1}$ Mpc$^{-1}$ and $\sigma_{8} = 0.8149$ and generate the initial conditions with the code \textsc{music} \citep[][]{Hahn2011} where we use transfer function at redshift 124 computed by \textsc{camb} \citep[][]{Lewis1999}. We adopt a comoving magnetic field of $10^{-14}$ G at this redshift.   
We can use this as an effective benchmark for our model to demonstrate performance at a lower resolution but with greater statistical variety. \\
\\
The three selected data sets allow us to test the performance of \textsc{Virgo} for shock classification in the most common simulation setups in 3D simulations of galaxy clusters.
\textsc{ClustHD} and \textsc{ClustMHD} consist of single cluster objects, whereas \textsc{BoxMHD} is a simulation of a cosmological volume containing several cluster objects.
Therefore, our problem scales from unsupervised classification of an unknown number of target classes from a single cluster to the same for an unknown number of clusters.
We reduce the full data set of the simulation to only the parameters relevant for the actual shock surfaces. These are the spatial positions $\mathbf{x}$, the smoothing length $h$, 
the sonic Mach number $\mathcal{M}_\mathrm{s}$ detected by the shock finder and the shock normal vector $\hat{\mathbf{n}}_\mathrm{s}$ 
\citep[see][]{Beck2016a}.

%
%

 


\section{\textsc{Virgo} Model Pipeline}
\label{sec:virgo}

We propose a new analysis pipeline to solve the unsupervised classification of an unknown number of cosmological shock waves in a scalable, probabilistic and physically-motivated way. 
We separate our approach into four steps: \\
\\
\noindent
1) For data pre-processing, we remove data points above a conservative velocity threshold ($\mathcal{M}_s \leq 15$) and rescale the data set to a zero mean and unit variance.
Large-velocity particles are rarely a part of shock waves surfaces.
Our analysis only uses the particle position $\mathbf{x}$ and shock normal vector $\hat{\mathbf{n}}_s$.
Each particle therefore is a 6-dimensional vector $\mathbf{q} = (x_x, x_y, x_z, \hat{n}_{sx}, \hat{n}_{sy}, \hat{n}_{sz})^{\top}$. \\
\\
\noindent
2) The raw simulation output is noisy with non-shock wave particles and not centered, as is illustrated in Fig.~\ref{fig:raw250x_shocks}a and Fig.~\ref{fig:25x_box_shocks}a. 
We use an RBF kernel with the Nyström approximation on the particle positions $\mathbf{x}$ with default length scale $l = 2 \cdot \text{Var}(\mathbf{x}) = 2$.
We apply PCA on the resulting matrix and use a GMM in the feature space with expectation maximization to separate the actual cluster of shock waves from non-shock wave particles by only accepting the densest GMM component distribution.
The density is estimated by particle number and occupied volume.
This approach exploits the inherent local density changes of the problem by using a stationary kernel to separate non-shock wave particles from relevant shock wave particles.
The step does not reduce the data set through the Nyström approximation, as it works reliably for $m$ as low as $100$.
We rescale the denoised data again to zero mean and unit variance. \\
\\
\noindent
3) We construct a physically motivated composite kernel $k_{\text{V}}$ by adding two separate composite kernels made up of Matérn-$\frac{5}{2}$ kernels $k_{\text{M}}$ and linear kernels $k_{\text{L}}$
\begin{align}
    k_1 (\mathbf{q}, \mathbf{q}^\prime) &= k_{\text{M}}( \mathbf{x}, \mathbf{x}^\prime) \cdot k_{\text{L}} ( \mathbf{x}, \mathbf{x}^\prime) \\
    k_2 (\mathbf{q}, \mathbf{q}^\prime) &= k_{\text{M}}( \mathbf{x}, \mathbf{x}^\prime) \cdot k_{\text{L}} ( \hat{\mathbf{n}}_s,\hat{\mathbf{n}}_s^\prime) \\
    k_{\text{V}} (\mathbf{q}, \mathbf{q}^\prime) &= k_1 (\mathbf{q}, \mathbf{q}^\prime) + k_2 (\mathbf{q}, \mathbf{q}^\prime) .
\end{align}
The first kernel $k_1$ creates a non-stationary kernel for spatial information, whereas the second kernel $k_2$ combines local spatial information with shock normal directions of the particles.
We combine $k_{\text{V}}$ with the Nyström approximation but increase $m$, accepting a reduction of the data set for computational limitations, and finally apply PCA again.
The resulting feature space enables separation with a fixed linking length $\beta$ FoF algorithm.
We estimate $\beta$ with the average n-next-neighbor distance in the resulting feature space.
We automatically label every unclassified particle as non-shock wave particle. \\
\\
\noindent
4) We use this now labeled subset to train an SV-DKL classifier with a setup as described in Sec.~\ref{subsec:svdkl} and \ref{sec:appl}. 
The deep kernel is adjusted to our problem regarding the accuracy and a minimal number of parameters and is a fully connected NN with ReLU activation functions after each layer.
We use an RBF kernel for the additive base kernels.
The inducing points and training parameters are set up as in \citep{svdkl_wilson2016} and \citep{denseNN_huang2017}, unless stated otherwise in Sec.~\ref{sec:appl}.
We use a standard softmax likelihood.
With the non-euclidean similarity metric of the deep kernel, we can do long-range extrapolation through our input space with local density changes.
This benefit implies a locally adaptable similarity metric required for robust classification.
The SV-DKL framework allows us to achieve fast inference and good scalability, as we are not limited by the size of the data set.\\
\\
Our approach is distinctly scalable, as the proposed pipeline succeeds for data sets of extreme sizes: We can downsize the data set at each step only to recover full resolution with the SV-DKL at the end.
We collect our analysis in a Python software package to be available for future analysis, called \textsc{Virgo} (\textbf{V}ariational \textbf{I}nference package for unsupe\textbf{R}vised classification of (inter-)\textbf{G}alactic sh\textbf{O}ck waves)\footnote{The source code is available at \href{https://github.com/maxlampe/virgo}{https://github.com/maxlampe/virgo}}.
The package utilizes already implemented features of PyTorch \citep{NEURIPS2019_9015}, GPyTorch \citep{gpytorch_gardner2018}, scikit-learn \citep{scikit_2011} and pyfof \citep{pyfof}.
\textsc{Virgo} has many high-level features for usability, such as a central data class to interact with other\textsc{Virgo} classes.
Next to the used models and kernels, we also added tools (plotting, cluster cleaning, result sorting, IO features, and others). 


\section{Experiments}
\label{sec:appl}

To test our pipeline, we evaluate on the described data sets from Sec.~\ref{sec:datasets}.
The generated data sets are the most straightforward data sets available, as this is a complex and previously unsolved problem.
As there exist no labeled data sets, it is hard to evaluate the factual accuracy of our pipeline.
However, we can verify the goodness of the overall results visually and evaluate our classifier from step 4) in Sec.~\ref{sec:virgo}.
The results are judged by the coherence of the shock wave surface classification and the removal of obvious non-shock particles.
In our studies on all data sets, we can observe that other approaches either over- or under-segment the shock waves, including all mentioned methods in \ref{sec:background} on their own. 
Some of these studies are presented in the ~\ref{sec:others}. 
The results for the remaining data sets for all steps are also schon in the \ref{sec:appendix}.
We optimized all parameters in steps 1) to 3) to run all experiments on a Linux machine with two $2.4$ GHz CPU cores and $8$ GB RAM to highlight its efficiency. However, step 4) required a GPU with 16 GB RAM.

\subsection{Evaluation on \textsc{ClustHD} and \textsc{ClustMHD} Data Sets}
\label{sec:appl_clust}

\begin{figure*}
    \centering
    \includegraphics[width=\textwidth]{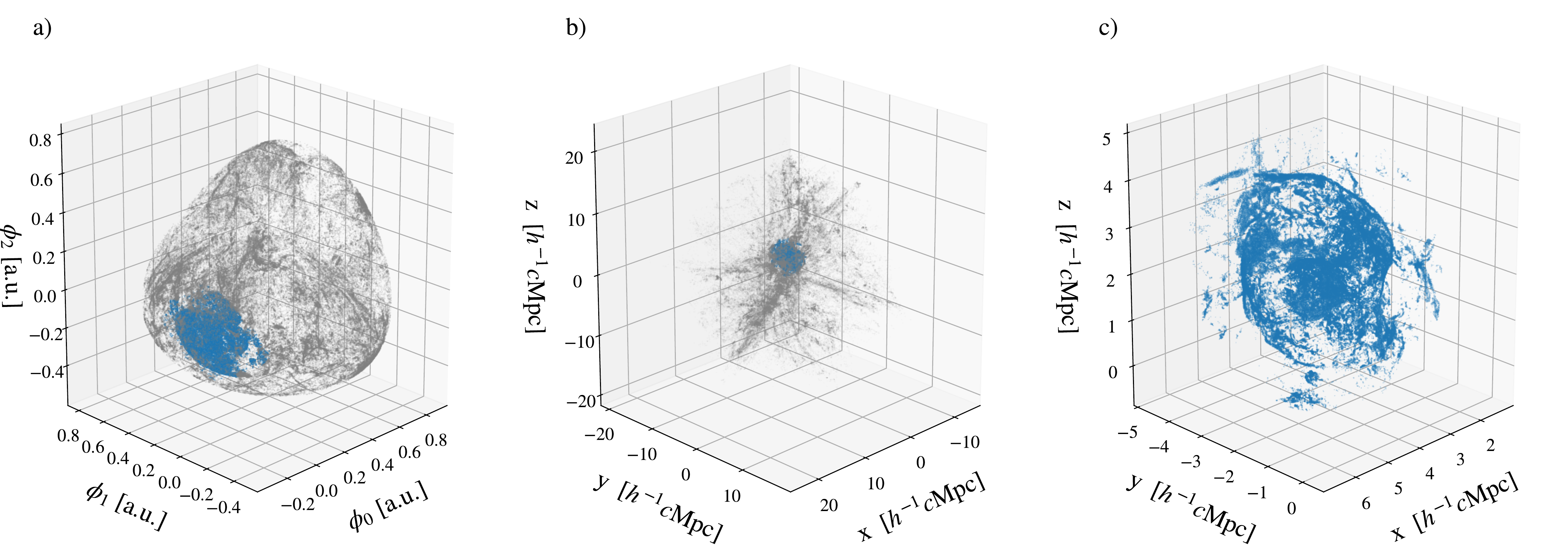}
    \caption{We show the denoising process of Sec.~\ref{sec:virgo} step 2) and its impact on the $\textsc{ClustHD}_2$ data set from Fig.~\ref{fig:raw250x_shocks}. \textit{a)} GMM fitted in kernel-PCA space with low-density noise component (gray) and high-density cluster component (blue). We only plot the first three kernel space dimensions. \textit{b)} Labeled data from \textit{a)}, but shown in physical space. \textit{c)} Resulting denoised data set for further analysis.}
    \label{fig:denoising}
\end{figure*}

\begin{figure*}
    \centering
    \includegraphics[width=\textwidth]{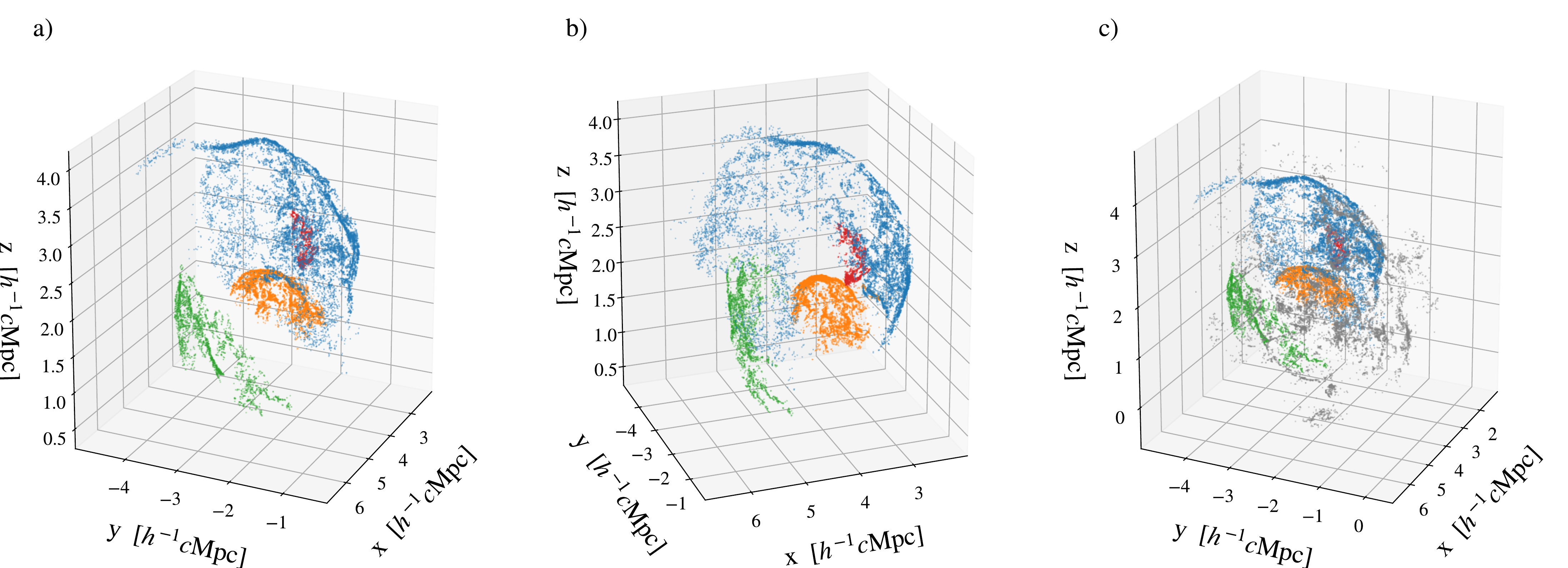}
    \caption{We show the labeled subset of $\textsc{ClustHD}_2$ after the denoising process. The clear separation of the shock wave surfaces from non-shock wave particles (gray) is visible. Also, the estimated error of labeling shock surface particles as non-shock wave particles is visibly negligible. \textit{a)} Labeled subset with the FoF algorithm of step 3) in Sec.~\ref{sec:virgo}. \textit{b)} same as a), but rotated by 45$^\circ$. \textit{c)} Same as a), but with non-shock wave particles plotted too.}
    \label{fig:labeled250hd_790}
\end{figure*}

\begin{figure*}
    \centering
    \includegraphics[width=\textwidth]{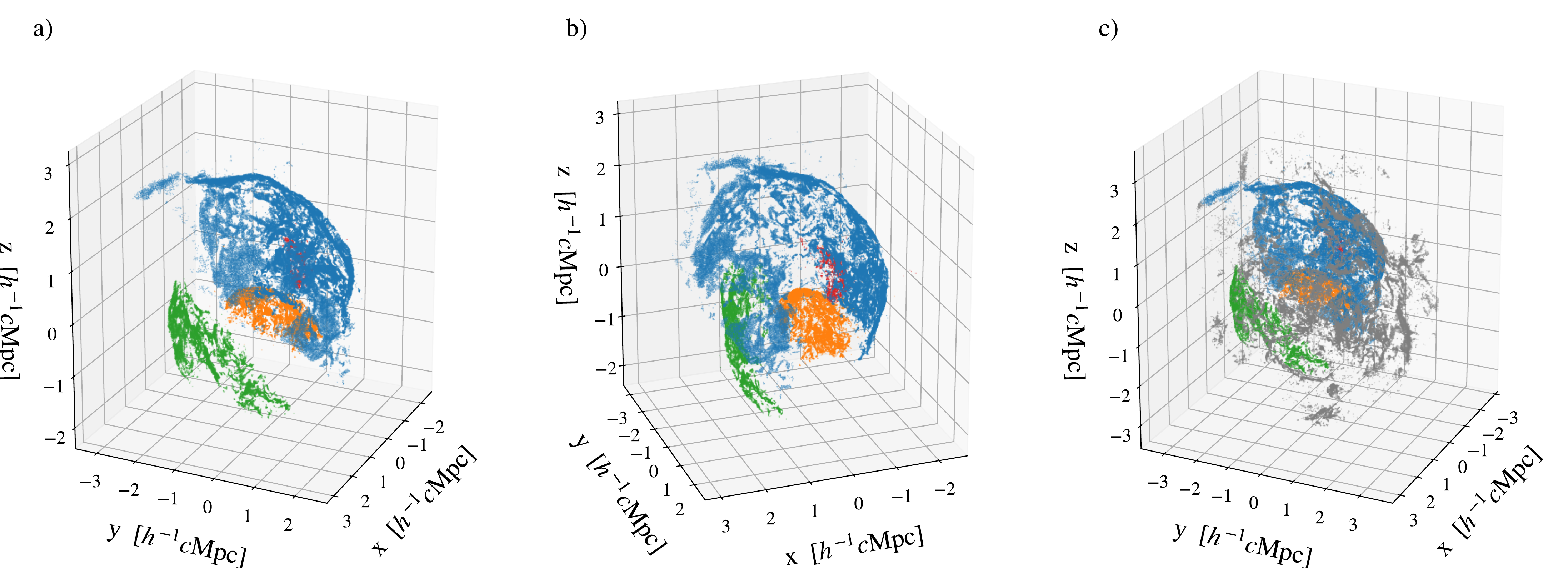}
    \caption{We show the full resolution reconstruction of $\textsc{ClustHD}_2$ with the SV-DKL classification. 
    This result is the final \textsc{Virgo} output for the raw input in Fig.~\ref{fig:raw250x_shocks} and allows scalability to large data sets.
    We state the accuracy of the recovery classification in Tab.~\ref{tab:comp_res}.
    \textit{a)} Labeled data set with SV-DKL classifier of step 4) in Sec.~\ref{sec:virgo}. \textit{b)} same as a), but rotated by 45$^\circ$. \textit{c)} Same as a), but with non-shock wave particles.}
    \label{fig:deep_790}
\end{figure*}

First, we evaluate one single cluster data set where only one shock wave structure is present within many non-shock wave particles.
We list any distinct parameters for our pipeline and its training:
For the denoising step 2), we use an RBF kernel with the Nyström approximation $m = 100$, PCA with $k = 5$ components, and GMM with $2$ or $5$ components, depending on the data set.
The physically motivated kernel $k_V$ of step 3) is used with Nyström approximation $m = 500$, PCA with $k = 6$ components, and we estimate the FoF linking length $\beta$ as the average 20-next-neighbor distance.
The deep kernel NN is set up with six input features, $n_h = 1$ hidden layer of size 20, $n_f = 10$ output features, and without pre-training.
We trained the SV-DKL of step 4) over 20 epochs, with a batch size of 1024, a grid of 64 inducing points, an overall learning rate of $\alpha = 0.1$, a learning rate scheduler decreasing it by $0.1$ after 50 and 75\% of the training time, decreased learning rate for the Gaussian process parameters $\alpha_{GP} = \alpha / 100$, L2 regularization with weight decay of $10^{-4}$, and an ADAM optimizer ($\beta_1 = 0.9$, $\beta_2 = 0.999$) \citep{kingma2014adam}.
We employed an 80/10/10 spit between training, validation and test sets. \\
\\
We show the denoising and centering process representative for $\textsc{ClustHD}_2$ in Fig.~\ref{fig:denoising}.
Our approach accurately separates the dense cluster region from the general simulation output of shocked particles.
Should more structures be present in the general output, we increase the number of GMM components and obtain reliable results. 
In doing so, we achieve good separation for all available and tested data sets, without exception. \\
\\
We classify the denoised result as described in step 3) with $k_V$, Nyström approximation, PCA and agglomerative clustering.
The resulting classification is illustrated for one data sets in Fig.~\ref{fig:labeled250hd_790}.
The figures show the clear separation and classification of the shock surfaces without over-segmentation.
Any non-shock-wave particles are labeled as noise with a visibly low error rate.
However, this step reduces the data set size to $\mathcal{O}(10^{4})$ from the original $\mathcal{O}(10^{6})$ after denoising.
We recover full resolution with the SV-DKL classifier trained on the labeled subset.
Fig.~\ref{fig:deep_790} shows the reconstructed and labeled full data set representative of $\textsc{ClustHD}_2$.
The complex morphology of the shock waves and its substructures are fully restored. \\
\\
To compare our classification, we benchmark the SV-DKL against a k-nearest-neighbor (k-NN, k = 10) classifier and a fully connected NN.
We construct the NN similarly to the deep kernel NN, i.e., one hidden layer of size 20, six input features, and $n$ target classes as output with ReLU activation functions.
We present the results in Tab.~\ref{tab:comp_res}.
Next to its other benefits, the SV-DKL outperforms the other methods in terms of accuracy.
The authors of \citep{dkl_wilson2016, svdkl_wilson2016} further illustrated the (SV)-DKL capabilities compared to other approaches.
We expect the SV-DKL to perform better with non-shock wave particles near shock wave surfaces, due to the additive Gaussian processes.
However, k-NN achieves good accuracy regarding the labeled subset. 
Given the lower hardware requirements of k-NN, we recommend it as a valuable replacement, e.g., for online applications in large-scale simulations. \\
\\
In addition, \textsc{Virgo} shows signs of generalization, as we used the trained classifier from the labeled subset of $\textsc{ClustHD}_2$ on the full data set of $\textsc{ClustHD}_3$ and obtained approximately ideal results as well.
However, this requires the same amount of target shock wave classes for both data sets.
Nevertheless, we can use this for future work for better performance over several time steps by training the SV-DKL on several data sets with similar classes.
Furthermore, due to the SV-DKL, \textsc{Virgo} scales to data sets of any size, given that the underlying model assumptions and inducing points describe the data sets with all facets.
For more complex problems, we recommend expanding the expressiveness of the deep kernel with advanced architectures, such as DenseNet \citep{denseNN_huang2017}.
Overall, our proposed pipeline solves the outlined classification problem of cosmological shock waves and delivers robust results.
%
%
%
%

\subsection{Evaluation on \textsc{BoxMHD} Data Set}


\begin{table}
  \caption{Comparing average test accuracies on the labeled subsets of the data after step 3) in Sec.~\ref{sec:virgo} for different methods on different data sets for ten independent runs. Models are trained from scratch and data sets are re-shuffled. The added index for the data sets indicate a different time step of the simulation. Different time steps of the same simulation are quasi-independent data sets, as the morphologies and number of target classes change from one time step to the next.} 
  \label{tab:comp_res}
  \centering
  \begin{tabular}{c|ccc}
    \toprule
    Method   & $k$-NN   &   FC-NN   &  SV-DKL \\
    \midrule 
    \midrule
    \textsc{ClustHD}$_1$  & $97.10 \pm 0.34$  &  $95.33 \pm 1.32$ & $\boldsymbol{97.57 \pm 0.54}$ \\
    \textsc{ClustHD}$_2$  & $96.61 \pm 0.32$ & $95.51 \pm 0.84$ & $\boldsymbol{97.00 \pm 0.49}$    \\
    \textsc{ClustHD}$_3$  & $97.19 \pm 0.30$ & $96.63 \pm 0.41$ & $\boldsymbol{98.36 \pm 0.18}$  \\
    \textsc{ClustMHD}$_1$ & $96.57 \pm 0.39$ & $96.19 \pm 0.50$ &  $\boldsymbol{98.08 \pm 0.16}$ \\
    \textsc{BoxMHD}$_1$ & $96.69 \pm 0.48$ & $95.05 \pm 0.69$ & $\boldsymbol{98.02 \pm 0.37}$\\
    \bottomrule
  \end{tabular}
\end{table}

\begin{figure*}
    \centering
    \includegraphics[width=\textwidth]{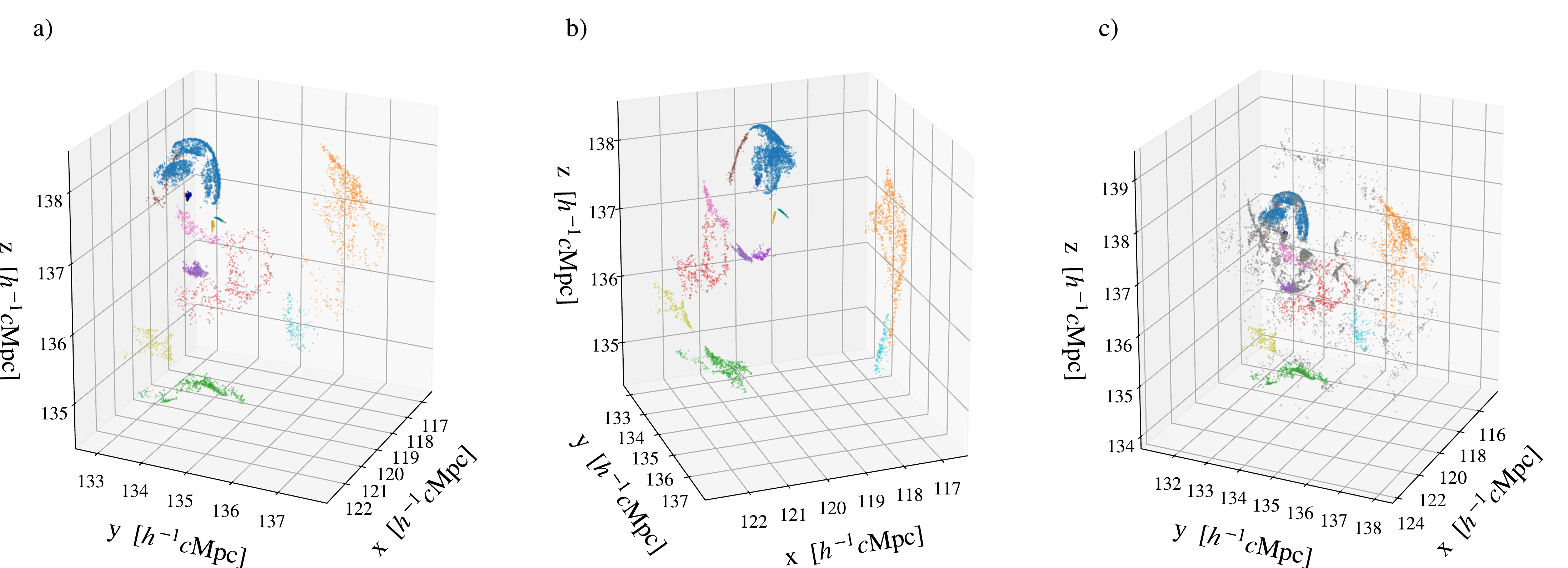}
    \caption{
    We show the labeled subset of $\textsc{BoxMHD}_1$ after the additional GMM selection and denoising process. The clear separation of the shock wave surfaces from non-shock wave particles (gray) is visible. Also, the estimated error of labeling shock surface particles as non-shock wave particles is negligible. We set a minimum class size during step 3). Hence, we ignore smaller structures and label them as non-shock wave particles. \textit{a)} Labeled subset with the FoF algorithm of step 3) in Sec.~\ref{sec:virgo}. \textit{b)} same as a), but rotated by 45$^\circ$. \textit{c)} Same as a), but with non-shock wave particles.
    }
    \label{fig:labeledbox0}
\end{figure*}

To test the extent of \textsc{Virgo}'s capabilities, we run the pipeline on the more complex $\textsc{BoxMHD}$ data set.
All other data sets consist of single cluster objects, whereas $\textsc{BoxMHD}$ is a simulation of a cosmological volume containing several cluster objects.
Therefore, we expand our problem of unsupervised classification of an unknown number of target classes from a single cluster to an unknown number of clusters.
We repeat step 2) for this data set twice, once with 35 GMM components, keeping the densest ten components, and then again on each remaining component with the parameters described above.
This additional step is required to single out dense objects and do single cluster shock wave analysis like in Sec.~\ref{sec:appl_clust}.
We illustrate the labeling of the subset from step 3) of the biggest substructure in $\textsc{BoxMHD}$ in Fig.~\ref{fig:labeledbox0}.
The scalability with the SV-DKL works as well as for the other data sets.
Some substructures are small enough for data size not to require the SV-DKL, as the Nyström approximation is not required.
Overall, \textsc{virgo} solves the outlined shock wave classification problem on all tested data sets.

\section{Discussion}
\label{sec:conclusion}

\subsection{Limitations}
\label{sec:appl_improv}

We determine the FoF linking length estimator from step 3), based on the average n-nearest-neighbor distance in kernel space, to be most prone to error.
As can be seen in Fig.~\ref{fig:labeledbox0}b, there is a small segment at the top (brown) that probably should belong to the bigger shock wave surface (blue).
However, our data sets are insufficient to construct a reliable estimator while avoiding overfitting.
Another issue to consider is error propagation.
Any labeling error in step 3) will be propagated by the SV-DKL.
The Gaussian process might correct minor errors for particles close to shock wave surfaces and edges, but this will not correct larger miss classifications.
To drive more accurate results and help mitigate the outlined issues, we propose manually creating labeled data sets by experts, creating synthetic data sets with labels, or generally looking at more simulation data.
Furthermore, we think it will be valuable to evaluate test accuracy in points close to the shock wave surfaces.
Future work should also verify the robustness of our chosen hyper-parameters on a wider set of simulated data, as this might pose a challenge for users. \\
\\
For applications to more complex data sets, such as $\textsc{BoxMHD}$, \textsc{Virgo} should be combined with a better structure finder.
The pipeline successfully classifies shock wave surfaces. However, we should expand the repeated GMM denoising approach from step 2) for cluster finding.
The GMM approach will struggle with extensive cosmological simulations, and there is a high risk of discarding smaller clusters.
Furthermore, \textsc{Virgo} currently lacks a criterion to determine whether it found shock waves.
We only tested the pipeline on data sets that contained shock wave clusters.\\
\\
Beyond the discussed approaches, we consciously did not use t-SNE \citep{tsne_maaten2008} due to a lack of interpretability in the resulting embedding space, and any of our attempts with latent variable models, e.g., with variational autoencoders \citep{kingma2013auto}, did not yield usable results.
However, we believe training the same DKL over different data sets with SV-DKL will yield an even more generalizable solution.
The DKL could be combined with PCA and $k$-NN to achieve better computational scalability, classification for an unknown number of shock waves in a cluster, and interpretability, as we could study the DKL for its mapping properties over several data sets.
The pre-trained DKL might even be able to replace or at least improve the subset labeling of step 2).

\subsection{Impact Statement}
\label{sec:appl_impact}

We demonstrated the capability of \textsc{Virgo} for capturing the irregular shapes of shock wave surfaces.
Accurately capturing these shock wave surfaces is a critical problem in astrophysics, specifically for particle acceleration at shock fronts which have to be identified \textit{on-the-fly} in large-scale numerical simulations of the Universe at non-negligible compute costs. 
For future work, we propose using \textsc{Virgo} to improve galaxy-cluster simulations by increasing the efficiency of particle injections.
Furthermore, we recommend applying \textsc{Virgo} to high-resolution simulations to reconstruct CR acceleration efficiency \citep[as used e.g. in][]{Kang2007} and study properties of shock waves such as Mach number distribution \citep[][]{Wittor2020}.
Future work could use \textsc{Virgo} to identify morphologically interesting shock structures, populate them with CRs spectra in post-processing and trace their time evolution using a Fokker-Planck solver \citep[e.g.][]{Donnert2014a, Wittor2017, Winner2019, Boess2022}.
As these solvers require a substantial amount of memory, applying them to a subset of the simulation is a promising compromise.
Beyond studying shock waves in clusters, we expect that \textsc{virgo} itself or by extent similar pipelines will enable machine learning applications to study supernovae remnants \citep{supern_janka2012} or general structure finding \citep[e.g.][]{Behroozi2013}. \\
\\
We believe our proposed pipeline will increase progress in cosmology simulation studies and lead to new applications in astrophysics. 
While our approach is unique and distinct, it is limited to specific synthetic data sets which are generated and therefore don't necessitate privacy or fairness considerations.
Hence, we think a broader impact discussion is not applicable.
Furthermore, given our specific data structure (spatial points and shock normal vectors), we see no possible applications outside natural sciences and for other purposes, even with malicious intent.
We cannot outline any societal impact beyond scientific interpretations of cosmological structure formation. 

\subsection{Conclusion}

We introduced a novel, physically motivated, and scalable pipeline. The unsupervised classification problem of cosmological shock waves was successfully solved for the first time.
We have only begun to explore the possibilities with physically motivated kernel methods combined with variational inference and deep kernel learning in astrophysics, including \textsc{Virgo} itself.
We hope our work inspires other astrophysics and physical sciences applications and that researchers will expand our approach to new problems with advanced kernel architectures or pipeline elements.

\subsection{Data Availability}

The data underlying this article will be shared on reasonable request to the corresponding author.

\section*{Acknowledgements}

ML, LMB and KD are acknowledging support by the Deutsche Forschungsgemeinschaft (DFG, German Research Foundation) under Germanys Excellence Strategy -- EXC-2094 -- 390783311. KD acknowledges support for the COMPLEX project from the European Research Council (ERC) under the European Union’s Horizon 2020 research and innovation program grant agreement ERC-2019-AdG 860744. The hydro-dynamical simulations were carried out at the Leibniz Supercomputer Center (LRZ) in Garching under the project pr86re and the CCA clusters ``rusty'' located in New York City as well as the cluster ``popeye'' located at the San Diego Supercomputing center (SDSC). UPS is supported by the Simons Foundation through a Flatiron Research Fellowship at the Center for Computational Astrophysics of the Flatiron Institute. The Flatiron Institute is supported by the Simons Foundation.


\bibliographystyle{mnras}
\bibliography{references} 

\appendix
\section{Appendix}
\label{sec:appendix}

This section provides additional plots to visualize further and underline the results and claims in Sec.~\ref{sec:appl}.
Furthermore, we hope to clarify the visual verification of results.
We present and discuss more results with \textsc{Virgo} as well as other attempts with other approaches to highlight the difficulty of the problem and distinct performance of \textsc{Virgo}.
For the latter section, we will only focus on the single cluster data set \textsc{ClustHD}$_3$, as more complex cases are far out of reach for other approaches.
All \textsc{Virgo} results are determined with the fully automatic pipeline for all data sets without fine-tuning.\\
\\
All assets, code, and data sets are cited and credited in the main paper.
We did not want to expand the main document with a large appendix for the submission and instead made an extensive supplementary document. 

\subsection{\textsc{Virgo} - Denoising and Centering}

\begin{figure*}
    \centering
    \includegraphics[width=\textwidth]{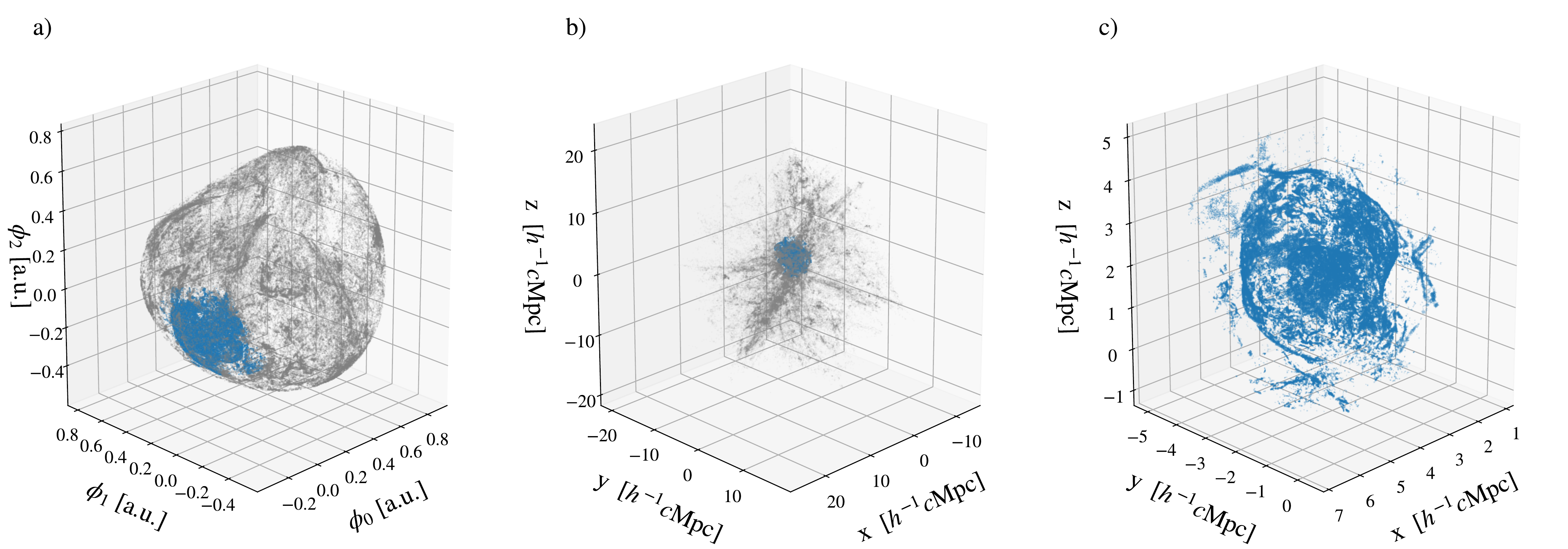}
    \caption{\textsc{Virgo} result - Denoising process for $\textsc{ClustHD}_3$ data set. \textit{a)} GMM fitted in kernel-PCA space with low-density noise component (gray) and high-density cluster component (blue). \textit{b)} Labeled data from \textit{a)}, but shown in physical space. \textit{c)} Resulting denoised data set.}
    \label{fig:denoise_800}
\end{figure*}

\begin{figure*}
    \centering
    \includegraphics[width=\textwidth]{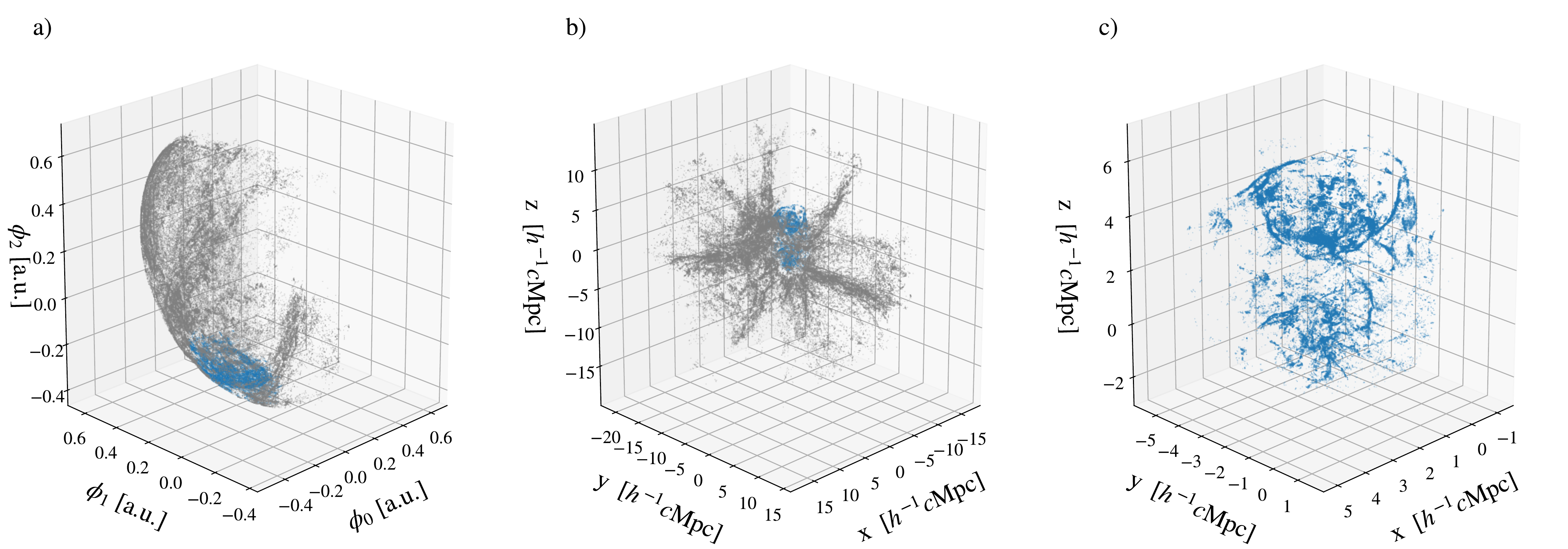}
    \caption{\textsc{Virgo} result - Denoising process for $\textsc{ClustMHD}_1$ data set. \textit{a)} GMM fitted in kernel-PCA space with low-density noise component (gray) and high-density cluster component (blue). \textit{b)} Labeled data from \textit{a)}, but shown in physical space. \textit{c)} Resulting denoised data set.}
    \label{fig:denoise_38}
\end{figure*}

We show the denoising process of Sec.~\ref{sec:virgo} step 2) from Fig.~\ref{fig:denoising} and its impact on other data sets.
The GMM is fitted in kernel-PCA space with low-density noise and high-density cluster component.
We only plot the first three kernel space dimensions and show the removal of non-shock wave particles before and after. \\
The Fig.~\ref{fig:denoise_800} and \ref{fig:denoise_38} show how \textsc{Virgo} ideally separates and centers on the cluster of shock waves from the enveloping cloud of non-shock wave particles.



\subsection{\textsc{Virgo} - Classification of Subset}

\begin{figure*}
    \centering
    \includegraphics[width=0.98\textwidth]{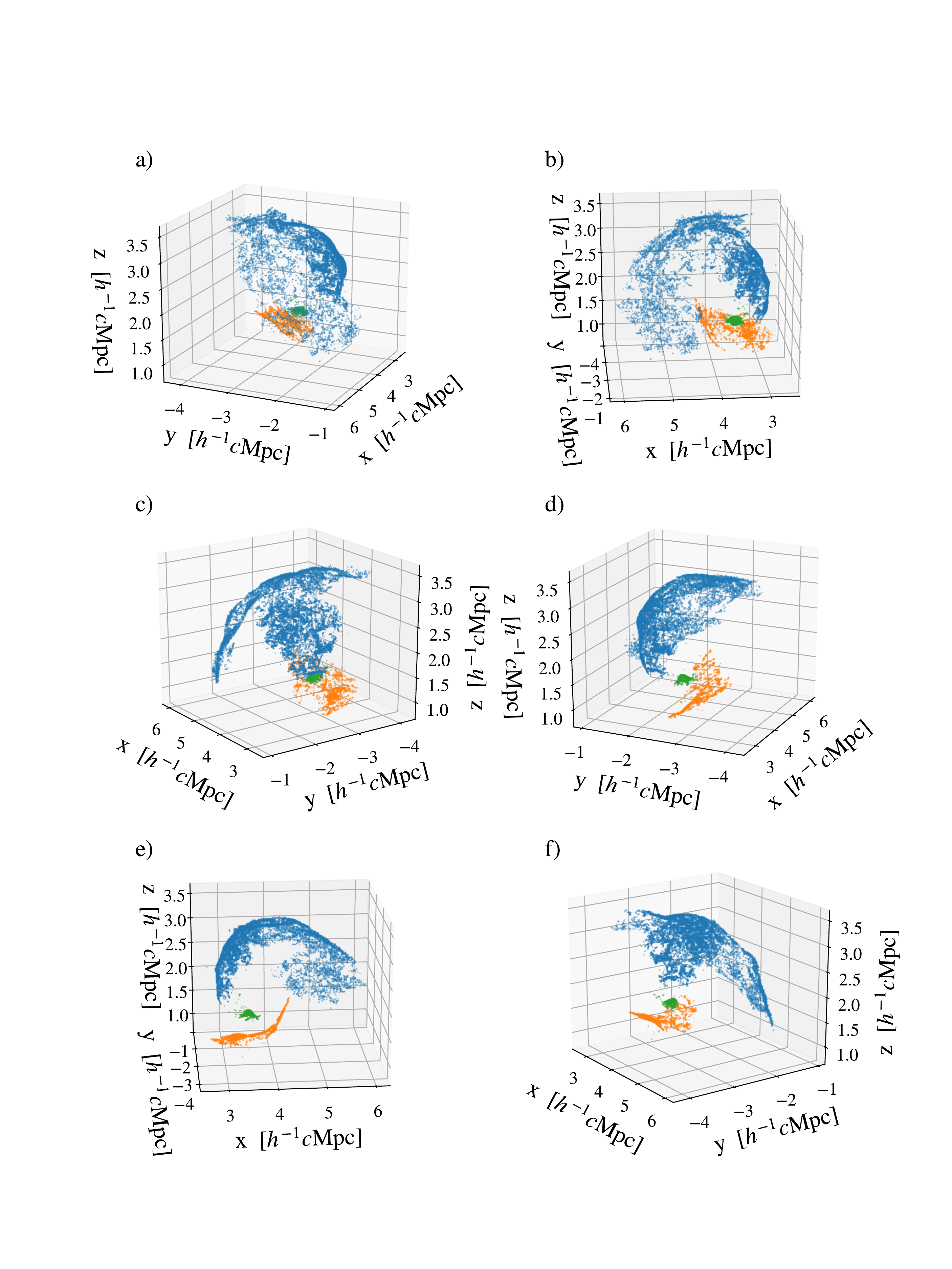}
    \caption{\textsc{Virgo} result - We show the labeled subset of $\textsc{ClustHD}_1$ after the denoising process. The classification is done as described in Sec.~\ref{sec:virgo} step 3). \textit{a)} - \textit{f)} Full rotation in 60$^\circ$ steps.}
    \label{fig:labeled_750}
\end{figure*}

\begin{figure*}
    \centering
    \includegraphics[width=0.98\textwidth]{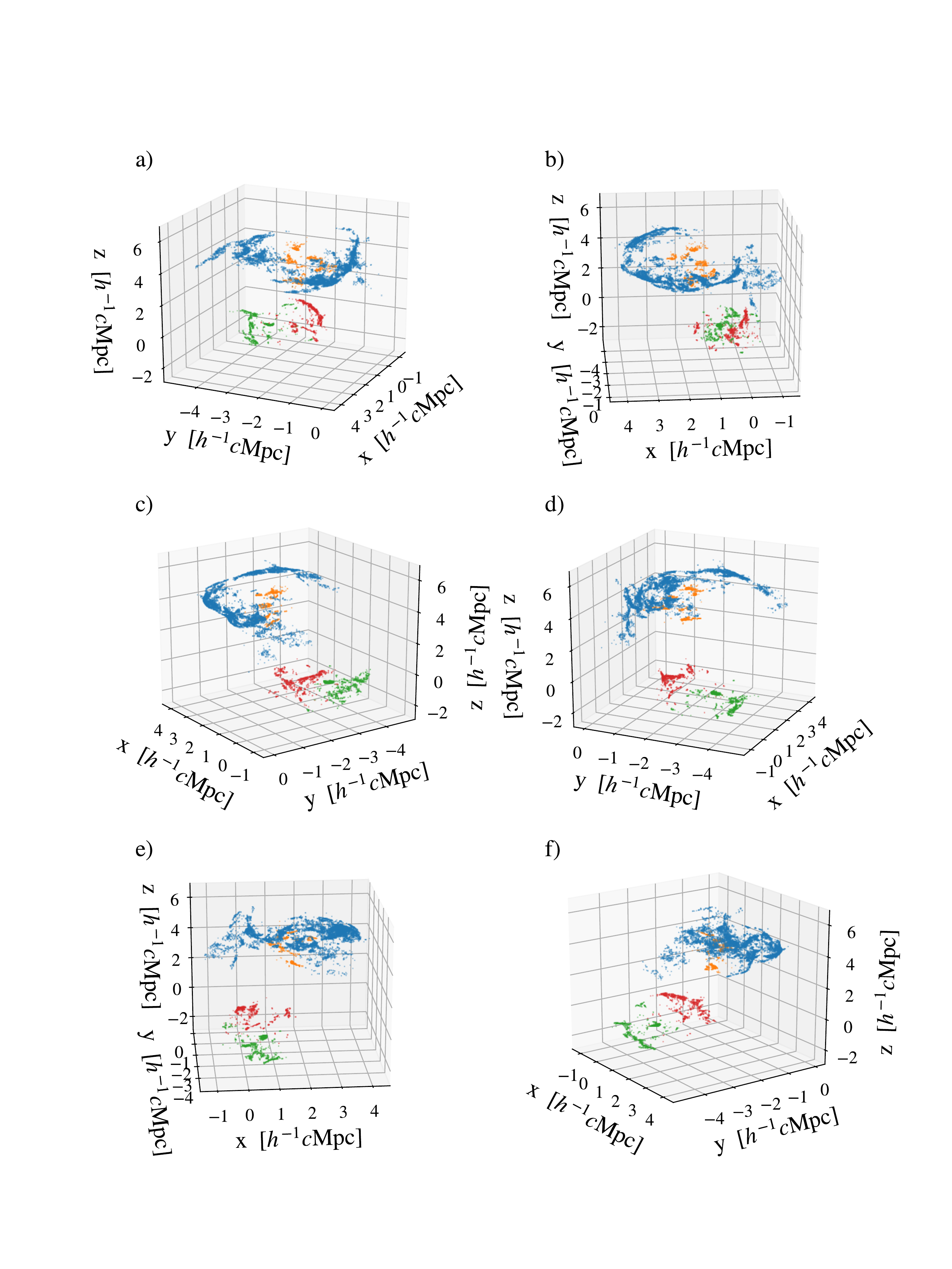}
    \caption{\textsc{Virgo} result - We show the labeled subset of $\textsc{ClustMHD}_1$ after the denoising process. The classification is done as described in Sec.~\ref{sec:virgo} step 3). \textit{a)} - \textit{f)} Full rotation in 60$^\circ$ steps.}
    \label{fig:labeled_38}
\end{figure*}

\begin{figure*}
    \centering
    \includegraphics[width=0.98\textwidth]{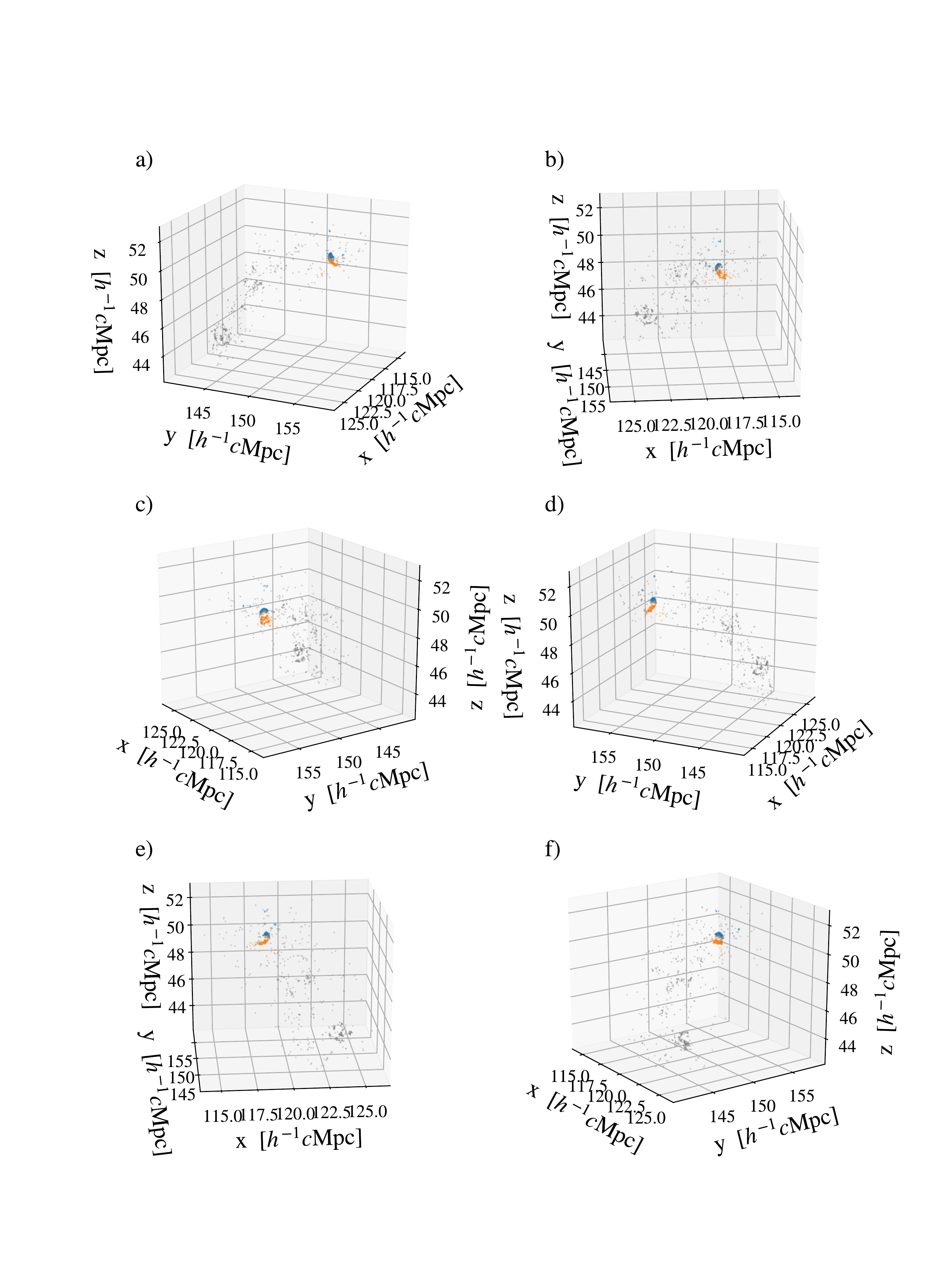}
    \caption{\textsc{Virgo} result - We show a labeled subset of $\textsc{BoxMHD}_1$ after the denoising process. The classification is done as described in Sec.~\ref{sec:virgo} step 3) and non-shock wave particles are plotted too. \textit{a)} - \textit{f)} Full rotation in 60$^\circ$ steps.}
    \label{fig:labeled_box4_noise}
\end{figure*}

Similar to Fig.~\ref{fig:labeled250hd_790}, we show the classification result of the denoised subsets of other data sets as well as an additional cluster from $\textsc{BoxMHD}$, $\textsc{BoxMHD}_2$. Each result is plotted for six different angles, once without classified non-shock wave particles.
The clear separation of the shock wave surfaces from non-shock wave particles is clearly visible, and the general quality of the classification in terms of coherent shock wave surfaces is verifiable. 
In addition, the estimated error of labeling shock surface particles as non-shock wave particles is visibly negligible. \\
Fig.~\ref{fig:labeled_750} shows an almost ideal result. There are some non-shock wave particles above the smallest shock wave (green).
The other two shock waves (blue, orange) are ideal.
Fig.~\ref{fig:labeled_38} shows how the most relevant and largest shock wave (blue) is ideally separated and classified.
However, smaller structures (all other colors) are hard to determine as shock wave structures, as they could also be a cluster of even smaller structures.
This differentiation and focus on larger structures when shock wave sizes vary is a limitation of \textsc{Virgo}. 
A criterion whether \textsc{Virgo} found a shock wave surface, as we discussed in Sec.~\ref{sec:conclusion}, would help to solve this problem for future work.
Fig.~\ref{fig:labeled_box4_noise} shows the classification of very small structures in a faint cloud of non-shock wave particles.
A few non-shock wave particles are misclassified as part of the shock wave structures.

\subsection{\textsc{Virgo} - SV-DKL Resolution Reconstruction}

\begin{figure*}
    \centering
    \includegraphics[width=0.98\textwidth]{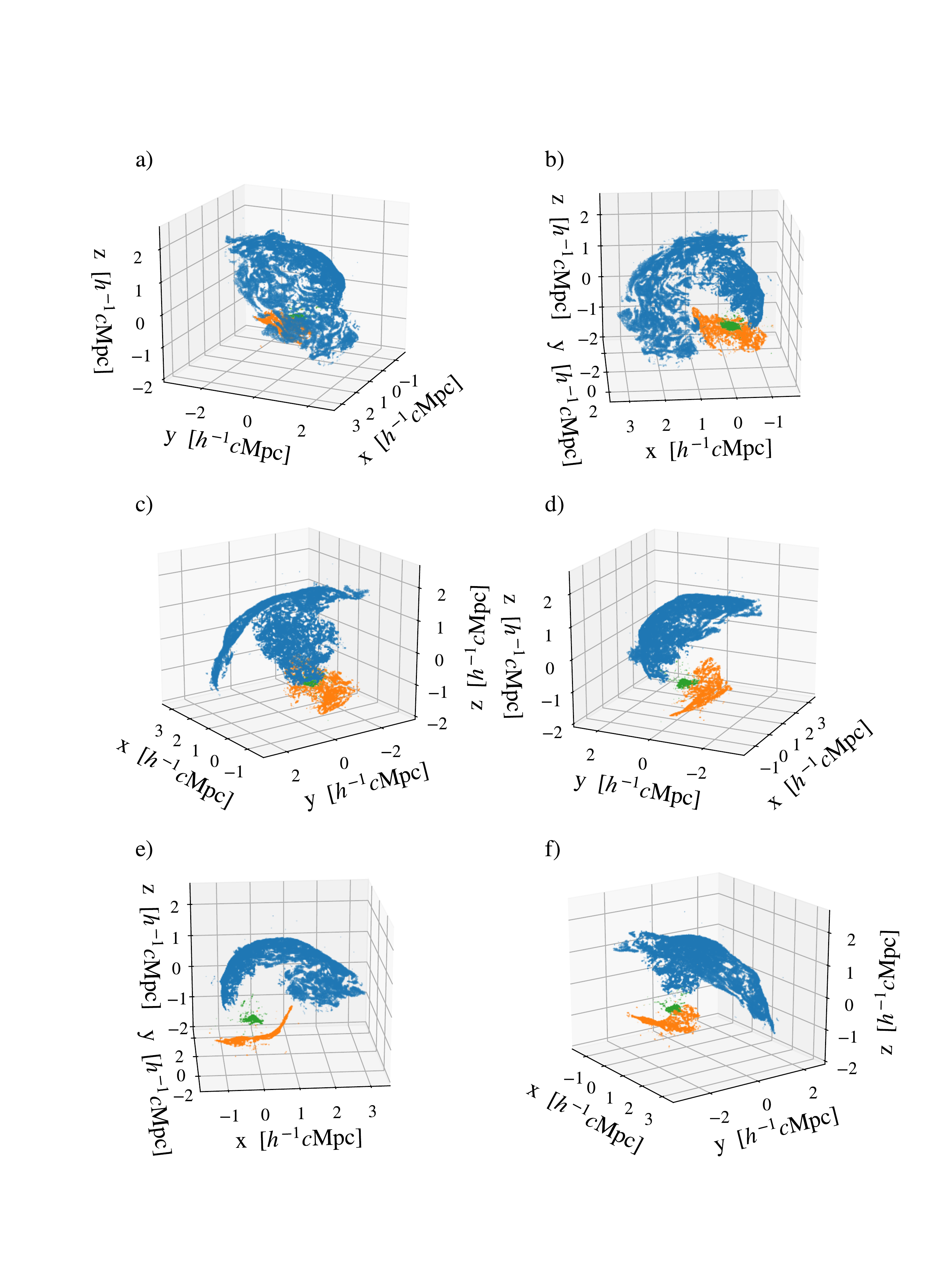}
    \caption{\textsc{Virgo} result - We show the full resolution reconstruction of $\textsc{ClustHD}_1$ after the denoising and subset labeling process. The classification is done as described in Sec.~\ref{sec:virgo} step 4) with SV-DKL. \textit{a)} - \textit{f)} Full rotation in 60$^\circ$ steps.}
    \label{fig:deep_750}
\end{figure*}

\begin{figure*}
    \centering
    \includegraphics[width=0.98\textwidth]{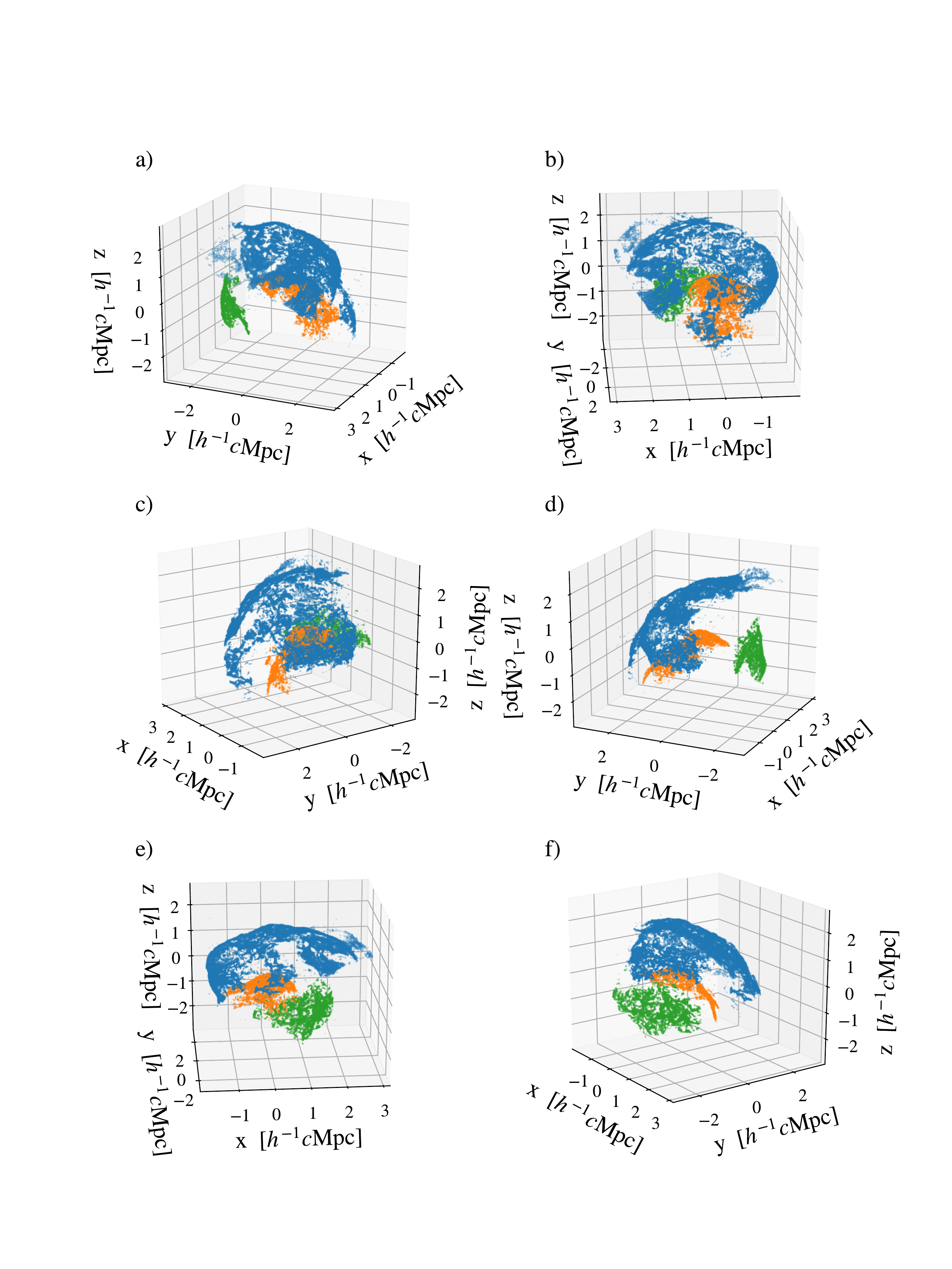}
    \caption{\textsc{Virgo} result - We show the full resolution reconstruction of $\textsc{ClustHD}_3$ after the denoising and subset labeling process. The classification is done as described in Sec.~\ref{sec:virgo} step 4) with SV-DKL. \textit{a)} - \textit{f)} Full rotation in 60$^\circ$ steps.}
    \label{fig:deep_800}
\end{figure*}

Similar to Fig.~\ref{fig:deep_790}, we show the full resolution reconstruction with the SV-DKL classifier. 
We trained it on the labeled subset from step 3) for each data set separately.
Each result is plotted for six different angles without classified non-shock wave particles due to illustrative purposes.
These results are the final \textsc{Virgo} outputs for the raw inputs and allow scalability to large data sets.
The separation of the shock wave surfaces from non-shock wave particles is clearly visible, and the general quality of the classification in terms of coherent shock wave surfaces is verifiable. \\
Fig.~\ref{fig:deep_750} shows an ideal result for the large shock wave surface (blue). 
The smaller shock waves (orange, green) are contaminated with a few non-shock wave particles, especially the smallest one (green).
The contamination is a propagated error from the previous labeling step and can be seen to some extent in Fig.~\ref{fig:labeled_750}.
%
%
Fig.~\ref{fig:deep_800} presents an ideal result with \textsc{Virgo}. All shock wave surfaces are correctly separated and classified. 
The SV-DKL classified all non-shock wave particles correctly as irrelevant.
%

\subsection{\textsc{Virgo} - SV-DKL Generalization Tests}

\begin{figure*}
    \centering
    \includegraphics[width=0.98\textwidth]{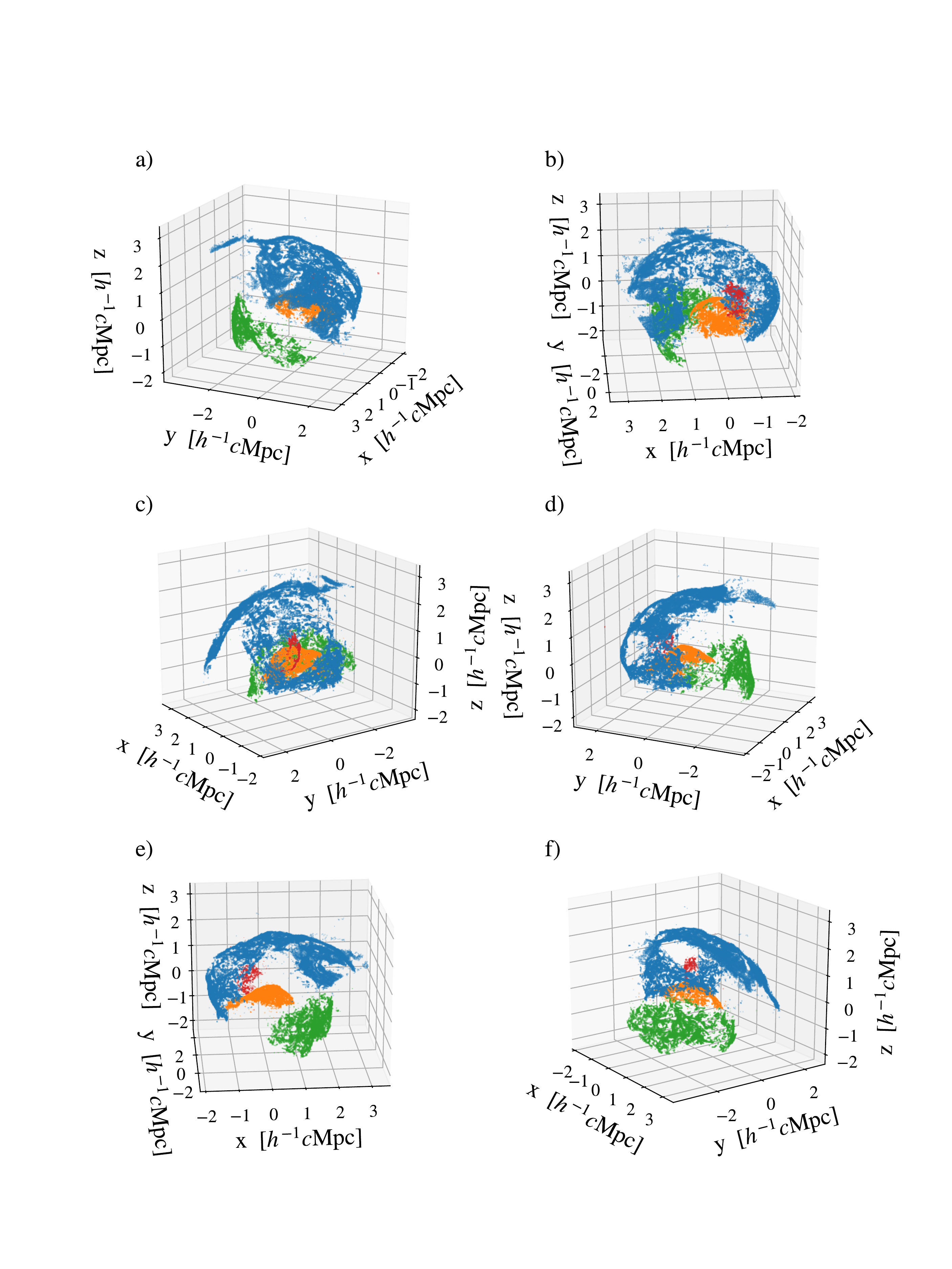}
    \caption{\textsc{Virgo} result - We show the generalization on the full resolution reconstruction of $\textsc{ClustHD}_3$. \textit{a)} - \textit{f)} Full rotation in 60$^\circ$ steps.}
    \label{fig:gen_800}
\end{figure*}

We claimed in Sec.~\ref{sec:appl} that the SV-DKL generalizes over different data sets, as long as the number of target classes is the same.
We tested the generalization on the $\textsc{ClustHD}$ data set by training the SV-DK on the labeled subset of time step $\textsc{ClustHD}_2$.
We show the achieved full resolution reconstructions of $\textsc{ClustHD}_3$, $\textsc{ClustHD}_4$ and $\textsc{ClustHD}_5$, which are all later time steps of the same simulation.
Different time steps mean changing shapes and shock wave propagation from one data set to another.\\
As can be seen in  
\ref{fig:gen_800}, \textsc{Virgo} provides good resolution reconstruction over different data sets, even though shock wave positions, surface densities and sizes change.
There is little to no contaminating noise on the results.

\subsection{Other approaches}
\label{sec:others}

\begin{figure*}
    \centering
    \includegraphics[width=0.98\textwidth]{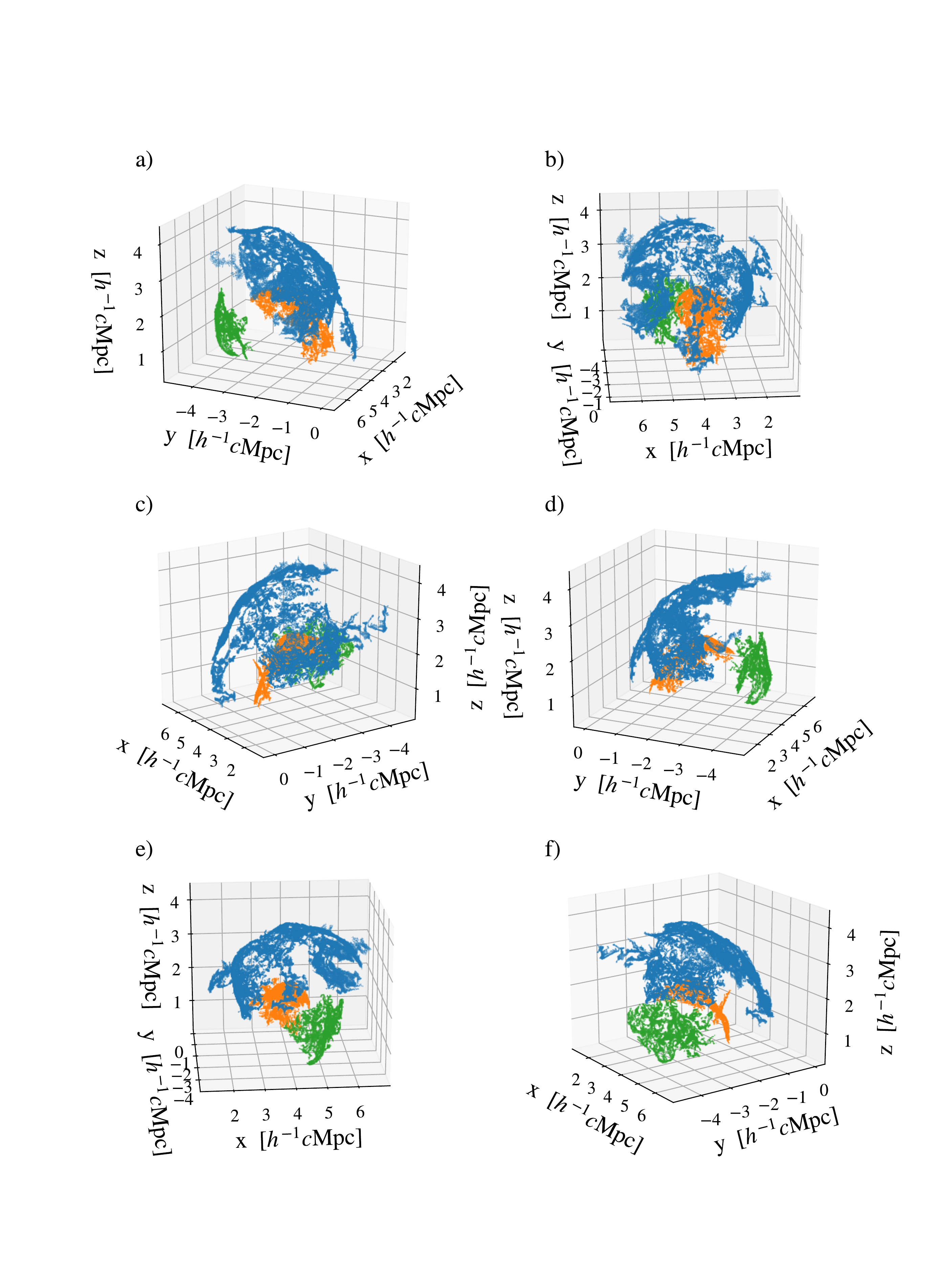}
    \caption{FoF on raw data set \textit{a)} - \textit{f)} Full rotation in 60$^\circ$ steps.}
    \label{fig:other_foffull}
\end{figure*}

\begin{figure*}
    \centering
    \includegraphics[width=0.98\textwidth]{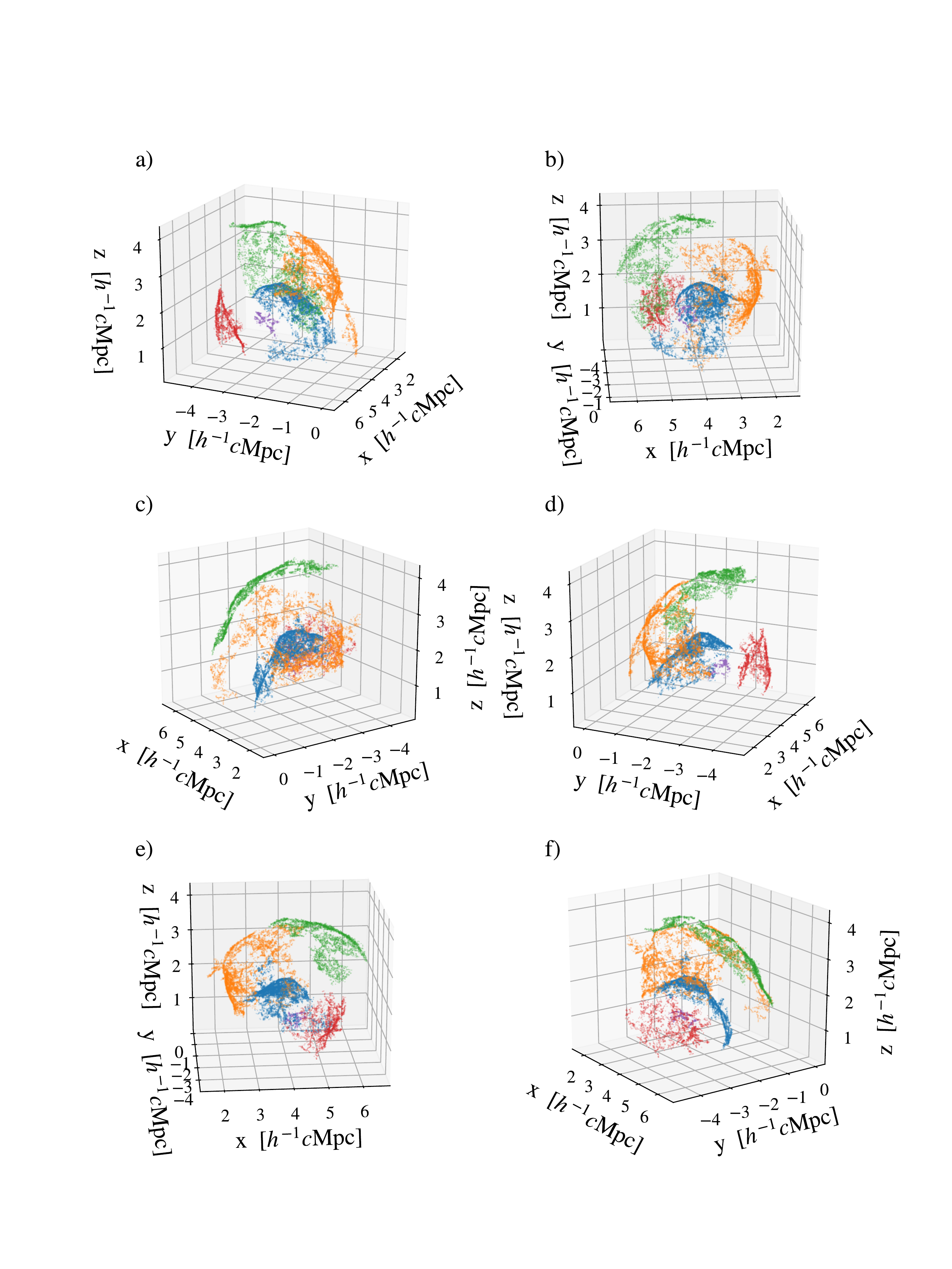}
    \caption{DBSCAN in space after denoising \textit{a)} - \textit{f)} Full rotation in 60$^\circ$ steps.}
    \label{fig:other_dbscan_denoised}
\end{figure*}

We present and discuss attempts with other approaches to highlight the difficulty of the problem and the distinct performance of\textsc{Virgo}.
For this section, we will only focus on the single cluster data set $\textsc{ClustHD}_3$, as more complex cases are far out of reach for other approaches.
Unlike with \textsc{Virgo} in previous sections, we manually fine-tune all results in this section to achieve the best possible results.
The used approaches are outlined and referenced in Sec.~\ref{sec:background}.\\
Fig.~\ref{fig:other_foffull} shows the application of a FoF algorithm with a fixed linking length on the full, raw data set.
Due to the high number of non-shock wave particles on the raw data set, any automatic linking length estimation fails, and we need to tune it manually ($\beta = 0.045$).
This result motivated the \textsc{Virgo} project.
There are problems while the full spatial resolution is intact and the shock wave objects are separated and classified.
The most significant shock wave (blue) has non-shock wave particles close above its surface, and another, separate structure clearly does not belong to the shock wave (Fig.~\ref{fig:other_foffull}b right, Fig.~\ref{fig:other_foffull}e left, Fig.~\ref{fig:other_foffull}f left).
Furthermore, the central shock wave structure (orange) has a separate structure not belonging to it wrongly classified (Fig.~\ref{fig:other_foffull}c bottom left, Fig.~\ref{fig:other_foffull}f bottom right).
Furthermore, the central shock wave structure (orange) is highly contaminated with non-shock wave particles above and even more so below its curved surface.
This problem is hard to illustrate with limited print capabilities, but it can be seen in the less dense following plots with similar issues.
The lack of automation as well as high contamination set the requirements for a solution such as \textsc{Virgo}.
Fig.~\ref{fig:other_dbscan_denoised} shows the application of DBSCAN on the denoised data set.
We manually set the distance parameter to $\epsilon = 0.11$ with a minimum density group size of 15.
Following previous points, even with manual tuning, any results with this approach are not scientifically usable.

\end{document}